\documentclass[12pt,tightenlines,eqsecnum,floats,aps,amsmath,amssymb,nofootinbib,prd]{revtex4}

\usepackage{amsmath,amssymb,amsfonts,latexsym}
\usepackage{graphicx}
\usepackage{enumerate} 
\usepackage{colordvi} 
\usepackage[mathscr]{eucal}
\usepackage{epstopdf}
\usepackage{colordvi}
\def\be{\begin{equation}}
\def\ee{\end{equation}}
\def\ba{\begin{eqnarray}}
\def\ea{\end{eqnarray}}

\def\R{\mathbb{R}}

\def\a{\mathfrak{a}}

\def\A{{\cal A}}
\def\Ab{\bar{\cal A}}

\def\H{{\cal H}}

\def\Hp{\H_{\rm phy}}

\def\Ab{{\bar \A}}
\def\h{\hat }
\def\lp{\ell_{\rm Pl}}
\def\M{M}
\def\SU(2){{\rm SU(2)}}
\def\su(2){{\rm su(2)}}

\def\su{{\rm su}}
\def\SU{{\rm SU}}

\def\lp{{\ell}_{\rm Pl}}

\def\e{{}^o\!e}
\def\w{{}^o\!\omega}
\def\grav{{\rm grav}}
\def\g{\gamma}

\newcommand{\ket}[1]{\ensuremath{|#1\rangle}}
\newcommand{\ip}[2]{{\langle#1\,|\,#2\rangle}}

\def\f{\frac}

\def\dd{\textrm{d}}
\def\ub{\underbar}
\def\ul{\underline}

\def\t{\tilde}
\def\p{\partial}

\usepackage[mathscr]{eucal}

\def\lp{\ell_{\rm Pl}}
\def\ps{\mathbf{\Gamma}}

\def\R{\mathbb{R}}
\def\M{M}

\def\SU(2){{\rm SU(2)}}
\def\su(2){{\rm su(2)}}
\def\U(1){{\rm U(1)}}
\def\A{{\cal A}}
\def\Ab{\bar{\cal A}}
\def\H{{\cal H}}

\begin{document}

\title{The Issue of the Beginning in Quantum Gravity}

\author{Abhay Ashtekar${}^{1,2}$}
\affiliation{Institute for Gravitational Physics and Geometry \\
Physics Department, Penn State, University Park, PA 16802-6300 \\
${}^2$ Institute for Theoretical Physics, University of Utrecht,
Princetonplein5, 3584 CC Utrecht, The Netherlands}

\maketitle

\section{INTRODUCTION}
\label{s1}

Treatise on Time, the Beginning and the End date back at least
twenty five centuries. Does the flow of time have an objective,
universal meaning beyond human perception? Or, it is fundamentally
only a convenient, and perhaps merely psychological, notion? Are
its properties tied to the specifics of observers such as their
location and state of motion? Did the physical universe have a
finite beginning or has it been evolving eternally? Leading
thinkers across cultures meditated on these issues and arrived at
definite but strikingly different answers. For example, in the
sixth century BCE, Gautama Buddha taught that `a period of time'
is a purely conventional notion, time and space exist only in
relation to our experience and the universe is eternal. In the
Christian thought, by contrast, the universe had a finite
beginning and there was debate whether time represents `movement'
of  bodies or if it flows only in the soul. In the fourth century
CE, St. Augustine held that time itself started with the world.

Founding fathers of modern Science from Galileo to Newton accepted
that God created the universe but Newton posited an absolute time
which is to run uniformly from the infinite past to the infinite
future. This paradigm became a dogma over centuries. Some
philosophers used it to argue that the universe itself \emph{had}
to be eternal. For, otherwise one could ask ``what was there
before?'' General relativity toppled this paradigm in one fell
swoop. Since the gravitational field is now encoded in space-time
geometry, geometry itself now becomes dynamical. The universe
could have had a finite beginning ---the big-bang--- at which not
only matter but space-time itself is `born'. General relativity
takes us back to St. Augustine's paradigm but in a detailed,
specific and mathematically precise form.  In semi-popular
articles and radio shows, relativists now like to emphasize that
the question ``what was there before?'' is rendered meaningless
because the notions of `before' and `after' refer to a space-time
geometry. We now have a new paradigm, a new dogma: In the
Beginning there was the big bang.

But general relativity is incomplete. For, it ignores quantum
effects entirely. Over the last century, we have learned that
these effects become important in the physics of the small and
should in fact dominate in parts of the universe where matter
densities become enormous. So, there is no reason to trust the
predictions of general relativity near space-time singularities.
The classical physics of general relativity does come to a  halt
at the big-bang. But applying general relativity near a
singularities is an extrapolation which has no justification
whatsoever. We need a theory that incorporates not only the
dynamical nature of geometry but also the ramifications of quantum
physics. Does the `correct' or `true' physics stop at the big-bang
also in quantum gravity? Or, is there yet another paradigm shift
waiting in the wings?

The goal of this article is to present an up to date summary of
the status of these age-old issues within loop quantum gravity
(LQG). Detailed calculations in simple cosmological models have
shown that the quantum nature of geometry does dominate physics
near the big bang, altering dynamics and drastically changing the
paradigm provided by general relativity. In particular, the
quantum space-time may be much larger than what general relativity
has us believe, whence the big bang may not, after all, be the
Beginning. Some of the mathematics underlying the main results is
subtle. However, I have made an attempt to also include a
descriptive summary of the viewpoint, ideas, constructions and
physical ramifications of the results.

\section{Loop Quantum Gravity}
\label{s2}

\subsection{Conceptual Issues}
\label{s2.1}

Remarkably, the necessity of a quantum theory of gravity was
pointed out by Einstein already in 1916. In a paper in the
Preussische Akademie Sitzungsberichte he wrote:
\begin{quote}
\textsl{Nevertheless, due to the inneratomic movement of
electrons, atoms would have to radiate not only electromagnetic
but also gravitational energy, if only in tiny amounts. As this is
hardly true in Nature, it appears that quantum theory would have
to modify not only Maxwellian electrodynamics but also the new
theory of gravitation.}
\end{quote}
Ninety years later, our understanding of the physical world is
vastly richer but a fully satisfactory unification of general
relativity with quantum physics still eludes us. Indeed, the
problem has now moved to the center-stage of fundamental physics.
(For a brief historical account of the evolution of ideas see,
e.g., \cite{aanjp}.)

A key reason why the issue is still open is the lack of
experimental data with direct bearing on quantum gravity. As a
result, research is necessarily driven by theoretical insights on
what the key issues are and what will `take care of itself' once
this core is understood.  As a consequence, there are distinct
starting points which seem natural. Such diversity is not unique
to this problem. However, for other fundamental forces we have had
clear-cut experiments to weed-out ideas which, in spite of their
theoretical appeal, fail to be realized in Nature. We do not have
this luxury in quantum gravity. But then, in absence of strong
experimental constraints, one would expect a rich variety of
internally consistent theories. Why is it then that we do not have
a single one? The reason, I believe, lies the deep conceptual
difference between the description of gravity in general
relativity and that of non-gravitational forces in other
fundamental theories. In those theories, space-time is given a
priori, serving as an inert background, a stage on which the drama
of evolution unfolds. General relativity, on the other hand, is
not only a theory of gravity, it is also a theory of space-time
structure. Indeed, as I remarked in section \ref{s1}, in general
relativity gravity is encoded in the very geometry of space-time.
Therefore, a quantum theory of gravity has to simultaneously bring
together \emph{gravity, geometry and the quantum}. This is a band
new adventure and our past experience with other forces can not
serve as a reliable guide.

LQG is an approach that attempts to face this challenge squarely
(for details, see, e.g., \cite{alrev,crbook,ttbook}). Recall that
Riemannian geometry provides the appropriate mathematical language
to formulate the physical, kinematical notions as well as the
final dynamical equations of any classical theory of relativistic
gravity. This role is now assumed by \textit{quantum} Riemannian
geometry. Thus, in LQG both matter and geometry are quantum
mechanical `from birth'.

In the classical domain, general relativity stands out as the best
available theory of gravity. Therefore, it is natural to ask:
\textit{Does quantum general relativity, coupled to suitable
matter} (or supergravity, its supersymmetric generalization)
\textit{exist as consistent theories non-perturbatively?} In
particle physics circles the answer is often assumed to be in the
negative, not because there is concrete evidence which rules out
this possibility, but because of the analogy to the theory of weak
interactions. There, one first had a 4-point interaction model due
to Fermi which works quite well at low energies but which fails to
be renormalizable. Progress occurred not by looking for
non-perturbative formulations of the Fermi model but by replacing
the model by the Glashow-Salam-Weinberg renormalizable theory of
electro-weak interactions, in which the 4-point interaction is
replaced by $W^\pm$ and $Z$ propagators. It is often assumed that
perturbative non-renormalizability of quantum general relativity
points in a similar direction. However this argument overlooks a
crucial and qualitatively new element of general relativity.
Perturbative treatments pre-suppose that space-time is a smooth
continuum \textit{at all scales} of interest to physics under
consideration. This assumption is safe for weak interactions. In
the gravitational case, on the other hand, the scale of interest
is \emph{the Planck length} and there is no physical basis to
pre-suppose that the continuum approximation should be valid down
to that scale. The failure of the standard perturbative treatments
may largely be due to this grossly incorrect assumption and a
non-perturbative treatment which correctly incorporates the
physical micro-structure of geometry may well be free of these
inconsistencies.

Are there any situations, outside LQG, where such physical
expectations are borne out by detailed mathematics? The answer is
in the affirmative. There exist quantum field theories (such as
the Gross-Neveu model in three dimensions) in which the standard
perturbation expansion is not renormalizable although the theory
is \emph{exactly soluble}! Failure of the standard perturbation
expansion can occur because one insists on perturbing around the
trivial, Gaussian point rather than the more physical non-trivial
fixed point of the renormalization group flow. Interestingly,
thanks to the recent work by Lauscher, Reuter, Percacci, Perini
and others, there is now growing evidence that situation may be
similar with general relativity (see \cite{lr} and references
therein). Impressive calculations have shown that pure Einstein
theory may also admit a non-trivial fixed point. Furthermore, the
requirement that the fixed point should continue to exist in
presence of matter constrains the couplings in physically
interesting ways \cite{pp}.

Let me conclude this discussion with an important caveat. Suppose
one manages to establish that non-perturbative quantum general
relativity (or, supergravity) does exist as a mathematically
consistent theory. Still, there is no a priori reason to assume
that the result would be the `final' theory of all known physics.
In particular, as is the case with classical general relativity,
while requirements of background independence and general
covariance do restrict the form of interactions between gravity
and matter fields and among matter fields themselves, the theory
would not have a built-in principle which \textit{determines}
these interactions. Put differently, such a theory would not be a
satisfactory candidate for unification of all known forces.
However, just as general relativity has had powerful implications
in spite of this limitation in the classical domain, quantum
general relativity should have qualitatively new predictions,
pushing further the existing frontiers of physics. In section
\ref{s3} we will see an illustration of this possibility.

\subsection{Salient features}
\label{s2.2}

Detailed as well as short and semi-qualitative reviews of LQG have
recently appeared in the literature (see, e.g., \cite{alrev} and
\cite{aanjp} respectively). Therefore, here I will only summarize
the key features of the theory that are used in section \ref{s3}.

The starting point of LQG is a Hamiltonian formulation of general
relativity based on spin connections \cite{aa}. Since all other
basic forces of nature are also described by theories of
connections, this formulation naturally leads to an unification of
all four fundamental forces at a \emph{kinematical} level.
Specifically, the phase space of general relativity is the same as
that of a Yang-Mills theory. The difference lies in dynamics:
whereas in the standard Yang-Mills theory the Minkowski metric
features prominently in the definition of the Hamiltonian, there
are no background fields whatsoever once gravity is switched on.

Let us focus on the gravitational sector of the theory. Then, the
phase space $\ps_\grav$ consists of canonically conjugate pairs
$(A_a^i, P_{a}^i)$, where $A_a^i$ is a connection on a 3-manifold
$\M$ and $P_{a}^i$ a vector density of weight one, both of which
take values in the Lie-algebra $\su(2)$. The connection $A$
enables one to parallel transport chiral spinors (such as the left
handed fermions of the standard electro-weak model) along curves
in $\M$. Its curvature is directly related to the electric and
magnetic parts of the space-time \emph{Riemann tensor}. $P^a_i$
plays a double role. Being the momentum canonically conjugate to
$A$, it is analogous to the Yang-Mills electric field. In
addition, $E^a_i := 8\pi G\g P^a_i$, has the interpretation of a
frame or an orthonormal triad (with density weight 1) on $\M$,
where $\g$ is the `Barbero-Immirzi parameter' representing a
quantization ambiguity. Each triad $E^a_i$ determines a positive
definite `spatial' 3-metric $q_{ab}$, and hence the Riemannian
geometry of $\M$. This dual role of $P$ is a reflection of the
fact that now $\SU(2)$ is the (double cover of the) group of
rotations of the orthonormal spatial triads on $\M$ itself rather
than of rotations in an `internal' space associated with $\M$.

To pass to quantum theory, one first constructs an algebra of
`elementary' functions on $\ps_\grav$ (analogous to the phase
space functions $x$ and $p$ in the case of a particle) which are
to have unambiguous operator analogs. In LQG, the configuration
variables are the holonomies $h_e$ built from $A_a^i$ which enable
us to parallel transport chiral spinors along edges $e$ and fluxes
$E_{S,f}$ of `electric fields' or `triads' (smeared with test
fields $f$) across 2-surfaces $S$. These functions generate a
certain algebra $\a$ (analogous to the algebra generated by
operators $\widehat{\exp i\lambda x}$ and $\h{p}$ in quantum
mechanics). The first principal task is to find representations of
this algebra. In that representation, \emph{quantum} Riemannian
geometry can be probed through the traid operators
$\hat{E}_{S,f}$, which stem from classical orthonormal triads.
Quite surprisingly the requirement of diffeomorphism covariance on
$M$ suffices to single out a \emph{unique} representation of $\a$
\cite{lost,cf2}! This recent result is the quantum geometry analog
to the seminal results by Segal and others that characterized the
Fock vacuum in Minkowskian field theories. However, while that
result assumes not only Poincar\'e invariance of the vacuum but
also specific (namely free) dynamics, it is striking that the
present uniqueness theorems make no such restriction on dynamics.
The requirement that there be a diffeomorphism invariant state is
surprisingly strong and makes the `background independent' quantum
geometry framework surprisingly tight.

This unique representation was in fact introduced already in the
mid-nineties \cite{al,jb1,mm,rs1,jb2} and has been extensively
used in LQG since then. The underlying Hilbert space is given by
$\H = L^2(\Ab, \dd\mu_o)$ where $\Ab$ is a certain completion of
the classical configuration space $\A$ consisting of smooth
connections on $M$ and $\mu_o$ is a diffeomorphism invariant,
faithful, regular Borel measure on $\Ab$. The holonomy (or
configuration) operators $\hat{h}_e$ act just by multiplication.
The momentum operators $\hat{P}_{S,f}$ act as Lie-derivatives. In
the classical theory, by taking suitable derivatives in $M$ of
holonomies $h_e$ along arbitrary edges $e$, one can recover the
connection from which the holonomy is built. However, in the
quantum theory, the operators $\h{h}_e$ are discontinuous and
there is no operator $\h{A}$ corresponding to the connection
itself.

Key features of this representation which distinguish it from,
say, the standard Fock representation of the quantum Maxwell field
are the following. While the Fock representation of photons makes
a crucial use of the background Minkowski metric, the above
construction is manifestly `background independent'. Second, as
remarked above, the connection itself is not represented as an
operator (valued distribution). Holonomy operators, on the other
hand, are well-defined. Third, the `triads' or `electric field'
operators now have purely discrete eigenvalues. Given a surface
$S$ and a region $R$ one can express the area $A_S$ and volume
$V_R$ using the triads. Although they are non-polynomial functions
of triads, the operators $\h{A}_S$ and $\h{V}_R$ are well-defined
and also have discrete eigenvalues. By contrast, such functions of
electric fields can not be promoted to operators on the Fock
space. Finally, and most importantly, the Hilbert space $\H$ and
the associated holonomy and (smeared) triad operators only provide
a \emph{kinematical} framework ---the quantum analog of the full
phase space. Thus, while elements of the Fock space represent
physical states of photons, elements of $\H$ are \emph{not} the
physical states of LQG. Rather, like the classical phase space,
the kinematic setup provides a home for \emph{formulating} quantum
dynamics.

In the Hamiltonian framework, the dynamical content of any
background independent theory is contained in its constraints. In
quantum theory, the Hilbert space $\H$ and the holonomy and
(smeared) triad operators thereon provide the necessary tools to
write down quantum constraint operators. Physical states are
solutions to these quantum constraints. Thus, to complete the
program, one has to: i) obtain the expressions of the quantum
constraints; ii) solve the constraint equations; iii) construct
the physical Hilbert space from the solutions (e.g. by the group
averaging procedure); and iv) extract physics from this physical
sector (e.g., by analyzing the expectation values, fluctuations of
and correlations between Dirac observables). While strategies have
been developed ---particularly through Thiemann's `Master
constraint program' \cite{ttmc}--- to complete these steps,
important open issues remain in the full theory. However, as
section \ref{s3} illustrates, the program has been completed in
mini and midi superspace models, leading to surprising insights
and answers to some long-standing questions.

\section{APPLICATION: HOMOGENEOUS ISOTROPIC COSMOLOGY}
\label{s3}

There is long list of questions about the quantum nature of the
big bang. For example:
\begin{quote}
\noindent$\bullet$ How close to the Big Bang does a smooth
space-time of general relativity make sense? In particular, can
one show from first principles that this approximation is
valid at the onset of inflation?\\
$\bullet$ Is the Big-Bang singularity naturally resolved by
quantum gravity? Or, is some external input such as a new
principle or a boundary condition at the Big Bang essential?\\
$\bullet$ Is the quantum evolution across the `singularity'
deterministic? Since one needs a fully non-perturbative framework
to answer this question in the affirmative, in the Pre-Big-Bang
\cite{pbb1} and Ekpyrotic/Cyclic \cite{ekp1,ekp2} scenarios, for
example, so far the answer is in the negative.\\
$\bullet$ If the singularity is resolved, what is on the `other
side'? Is there just a `quantum foam', far removed from any
classical space-time, or, is there another large, classical
universe?
\end{quote}
\noindent For many years, these and related issues had been
generally relegated to the `wish list' of what one would like the
future, satisfactory quantum gravity theory to eventually address.
It seems likely that these issues can be met head-on only in a
background independent, non-perturbative approach. One such
candidate is LQG. Indeed, starting with the seminal work of
Bojowald some five years ago \cite{mb1}, notable progress has been
made in the context of symmetry reduced, minisuperspaces. Earlier
papers focussed only on singularity resolution. However, to
describe physics in detail, it is essential to construct the
physical Hilbert space and introduce interesting observables and
semi-classical states by completing the program outlined at the
end of the last section. These steps have been completed recently.
In this section, I will summarize the state of the art,
emphasizing these recent developments. (For a comprehensive review
of the older work see, e.g., \cite{mbrev}.)

Consider the spatially homogeneous, isotropic, $k\!\!=\!\!0$
cosmologies with a massless scalar field. It is instructive to
focus on this model because \emph{every} of its classical
solutions has a singularity. There are two possibilities: In one
the universe starts out at the big bang and expands, and in the
other it contracts into a big crunch. The question is if this
unavoidable classical singularity is naturally tamed by quantum
effects. This issue can be analyzed in the geometrodynamical
framework used in older quantum cosmology. Unfortunately, the
answer turns out to be in the negative. For example, if one begins
with a semi-classical state representing an expanding classical
universe at late times and evolves it back via the Wheeler DeWitt
equation, one finds that it just follows the classical trajectory
into the big bang singularity \cite{aps2,aps3}.

In loop quantum cosmology (LQC), the situation is very different
\cite{aps1,aps2,aps3}. This may seem surprising at first. For, the
system has only a finite number of degrees of freedom and von
Neumann's theorem assures us that, under appropriate assumptions,
the resulting quantum mechanics is unique. The only remaining
freedom is factor-ordering and this is generally insufficient to
lead to qualitatively different predictions. However, for reasons
I will now explain, LQC does turn out to be qualitatively
different from the Wheeler-DeWitt theory \cite{abl}.

Because of spatial homogeneity and isotropy, one can fix a
fiducial (flat) triad $\e^a_i$ and its dual co-triad $\w_a^i$. The
$\SU(2)$ gravitational spin connection $A_a^i$ used in LQG has
only one component $c$ which furthermore depends only on time;
$A_a^i = c\,\, \w_a^i$. Similarly, the triad $E^a_i$ (of density
weight 1) has a single component $p$;\, $E^a_i = p\,(\det \w)\,
\e^a_i$. $p$ is related to the scale factor $a$ via $a^2 = |p|$.
However, $p$ is not restricted to be positive; under $p
\rightarrow -p$ the metric remains unchanged but the spatial triad
flips the orientation. The pair $(c,p)$ is `canonically conjugate'
in the sense that the only non-zero Poisson bracket is given by:
\be \{c,\, p\} = \f{8\pi G \g}{3}\, ,\ee
where as before $\g$ is the Barbero-Immirzi parameter.

Since a precise quantum mechanical framework was not available for
full geometrodynamics, in the Wheeler-DeWitt quantum cosmology one
focused just on the reduced model, without the benefit of guidance
from the full theory. A major difference in Loop quantum cosmology
(LQC) is that although the symmetry reduced theory has only a
finite number of degrees of freedom, quantization is carried out
by closely mimicking the procedure used in \emph{full} LQG,
outlined in section \ref{s2}. Key differences between LQC and the
older Wheeler-DeWitt theory can be traced back to this fact.

Recall that in full LQG diffeomorphism invariance leads one to a
specific kinematical framework in which there are operators
$\h{h}_e$ representing holonomies and $\h{P}_{S,f}$ representing
(smeared) momenta but there is no operator(-valued distribution)
representing the connection $A$ itself \cite{lost,cf2}. In the
cosmological model now under consideration, it is sufficient to
evaluate holonomies along segments $\mu\,\e^a_i$ of straight lines
determined by the fiducial triad $\e^a_i$. These holonomies turn
out almost periodic functions of $c$, i.e. are of the form
$N_{(\mu)} (c):= \exp i\mu (c/2)$, where the word `almost' refers
to the fact that $\mu$ can be any real number. These functions
were studied exhaustively by the mathematician Harold Bohr, Niels'
brother. In quantum geometry, the $N_{(\mu)}$ are the LQC analogs
of the spin-network functions of full LQG.

In quantum theory, then, we are led to a representation in which
operators $\h{N}_{(\mu)}$ and $\h{p}$ are well-defined, but there
is \emph{no} operator corresponding to the connection component
$c$. This seems surprising because our experience with quantum
mechanics suggests that one should be able to obtain the operator
analog of $c$ by differentiating $\hat{N}_{(\mu)}$ with respect to
the parameter $\mu$. However, in the representation of the basic
quantum algebra that descends to LQC from full LQG, although the
$\h{N}_{(\mu)}$ provide a 1-parameter group of unitary
transformations, it fails to be weakly continuous in $\mu$.
Therefore one can not differentiate and obtain the operator analog
of $c$. In quantum mechanics, this would be analogous to having
well-defined (Weyl) operators corresponding to the classical
functions $\exp i\mu x$ but no operator $\h{x}$ corresponding to
$x$ itself. This violates one of the assumptions of the
von-Neumann uniqueness theorem. New representations then become
available which are \emph{inequivalent} to the standard
Schr\"odinger one. In quantum mechanics, these representations are
not of direct physical interest because we need the operator
$\h{x}$. In LQC, on the other hand, full LQG naturally leads us to
a new representation, i.e., to \emph{new quantum mechanics.} This
theory is inequivalent to the Wheeler-DeWitt type theory already
at a kinematical level. In the Wheeler-Dewitt theory, the
gravitational Hilbert space would be $L^2(\R, \dd c)$, operators
$\hat{c}$ would act by multiplication and $\hat{p}$ would be
represented by $-i\hbar \dd/\dd c$. In LQC the `quantum
configuration space' is different from the classical configuration
space: Just as we had to complete the space $\A$ of smooth
connections to the space $\Ab$ of generalized connections in LQG,
we are now led to consider a completion ---called the Bohr
compactification $\bar\R_{\rm Bohr}$---  of the `$c$-axis'. The
gravitational Hilbert space is now $L^2(\bar{\R}_{\rm Bohr},
\dd\mu_{\rm Bohr})$ \cite{abl} where $\dd\mu_{\rm Bohr}$ is the
LQC analog of the measure $\dd\mu_o$ selected by the uniqueness
results \cite{lost,cf2} in full LQG. The operators
$\hat{N}_{(\mu)}$ act by multiplication and $\hat{p}$ by
differentiation. However, there is no operator $\h{c}$. In spite
of these differences, in the semi-classical regime LQC is well
approximated by the Wheeler-DeWitt theory. However, important
differences manifest themselves at the Planck scale. These are the
hallmarks of quantum geometry \cite{alrev,mbrev}.

The new representation also leads to a qualitative difference in
the structure of the Hamiltonian constraint operator: the
gravitational part of the constraint is a \emph{difference}
operator, rather than a differential operator as in the
Wheeler-DeWitt theory. The derivation \cite{abl,aps2,aps3} can be
summarized briefly as follows. In the classical theory, the
gravitational part of the constraint is given by $\int d^3x\,
\epsilon^{ijk} e^{-1} E^a_i E^b_j F_{ab\, k}$ where $e = |\det
E|^{1/2}$ and $F_{ab}^k$ the curvature of the connection $A_a^i$.
The part $\epsilon^{ijk} e^{-1} E^a_i E^b_j$  of this operator
involving triads can be quantized \cite{mb1,abl} using a standard
procedure introduced by Thiemann in the full theory \cite{ttbook}.
However, since there is no operator corresponding to the
connection itself, one has to express $F_{ab}^k$ as a limit of the
holonomy around a loop divided by the area enclosed by the loop,
as the area shrinks to zero. Now, quantum geometry tells us that
the area operator has a minimum non-zero eigenvalue, $\Delta$, and
in the quantum theory it is natural to shrink the loop only till
it attains this minimum. There are two ways to implement this idea
in detail (see \cite{abl,aps2,aps3}). In both cases, it is the
existence of the `area gap' $\Delta$ that leads one to a
difference equation. So far, most of the LQC literature has used
the first method \cite{abl,aps2}. In the resulting theory, the
classical big-bang is replaced with a quantum bounce with a number
of desirable features. However, it also has one serious drawback:
at the bounce, matter density can be low even for physically
reasonable choices of quantum states. Thus, that theory predicts
certain departures from classical general relativity even in the
low curvature regime (for details, see \cite{aps2,aps3}). The
second  and more recently discovered method \cite{aps3} cures this
problem while retaining the physically appealing features of the
first and, furthermore, has a more direct motivation. For brevity,
therefore, I will confine myself only to the second method.

Let us represent states as functions $\Psi(v,\phi)$, where $\phi$
is the scalar field and the dimensionless real number $v$
represents geometry. Specifically, $|v|$ is the eigenvalue of the
operator $\hat{V}$ representing volume%
\footnote{In non-compact spatially homogeneous models, integrals
of physical interest over the full spatial manifold diverge.
Therefore, to obtain a consistent Hamiltonian description, one has
to introduce an elementary cell ${\cal V}$ and restrict all
integrals to ${\cal V}$ already in the classical theory. This is
necessary also in geometrodynamics. $\hat{V}$ is the volume
operator associated with ${\cal V}$.}
(essentially the cube of the scale factor):
\be \hat{V}\ket{v} =   K\,(\f{8\pi\g}{6})^{\f{3}{2}}\,\, |v|
\,\lp^3 \,\ket{v}\quad {\rm where}\quad K=
\f{3\sqrt{3\sqrt{3}}}{2\sqrt{2}}\, . \ee
Then, the LQC Hamiltonian constraint assumes the form:
\ba \label{hc3} \p^2_\phi \Psi(v,\phi)  &=& \nonumber  [B(v)]^{-1}
\, \left(C^+(v)\, \Psi(v+4,\phi) + C^o(v) \, \Psi(v,\phi)
+C^-(v)\, \Psi(v-4,\phi)\right)\\
&=:& - \Theta \,\Psi(v,\phi) ~ \ea
where the coefficients $C^\pm(v)$, $C^o(V)$ and $B(v)$ are given
by:
\ba C^+(v) &=& \nonumber \f{3\pi K G}{8} \, |v + 2| \,\,\,
\big| |v + 1| - |v +3|  \big|  \\
C^-(v) &=& \nonumber C^+(v - 4) \quad{\rm and}\quad C^o(v) = -
C^+(v) - C^-(v) \\
B(v) &=& \left(\f{3}{2}\right)^3 \, K\,\, |v| \, \bigg| |v +
1|^{1/3} - |v - 1|^{1/3} \bigg|^3  ~. \ea

Now, in each classical solution, $\phi$ is a globally monotonic
function of time and can therefore be taken as the dynamical
variable representing an \emph{internal} clock. In quantum theory
there is no space-time metric, even on-shell. But since the
quantum constraint (\ref{hc3}) dictates how $\Psi(v,\phi)$
`evolves' as $\phi$ changes, it is convenient to regard the
argument $\phi$ in $\Psi(v,\phi)$ as \emph{emergent time} and $v$
as the physical degree of freedom. A complete set of Dirac
observables is then provided by the constant of motion
$\h{p}_\phi$ and operators $\h{v}|_{\phi_o}$ determining the value
of $v$ at the `instant' $\phi=\phi_o$.

Physical states are the (suitably regular) solutions to Eq
(\ref{hc3}). The map $\h\Pi$ defined by $\h\Pi\, \Psi(v, \phi) =
\Psi(-v, \phi)$ corresponds just to the flip of orientation of the
spatial triad (under which geometry remains unchanged); $\h\Pi$ is
thus a large gauge transformation on the space of solutions to Eq.
(\ref{hc3}). One is therefore led to divide physical states into
sectors, each providing an irreducible, unitary representation of
this gauge symmetry. Physical considerations \cite{aps2,aps3}
imply that we should consider the symmetric sector, with
eigenvalue +1 of $\h{\Pi}$.

To endow this space with the structure of a Hilbert space, one can
proceed along one of two paths. In the first, one defines the
action of the Dirac observables on the space of suitably regular
solutions to the constraints and selects the inner product by
demanding that these operators be self-adjoint \cite{aabook}. A
more systematic procedure is the  `group averaging method'
\cite{dm}. The technical implementation \cite{aps2,aps3} of both
these procedures is greatly simplified by the fact that the
difference operator $\Theta$ on the right side of (\ref{hc3}) is
independent of $\phi$ and can be shown to be self-adjoint and
positive definite on the Hilbert space $L^2(\bar{\R}_{\rm Bohr},
B(v) \dd\mu_{\rm Bohr})$.

The final result can be summarized as follows. Since $\Theta$ is a
difference operator, the physical Hilbert space $\Hp$ has sectors
$\H_\epsilon$ which are superselected; $\Hp = \oplus_\epsilon
\H_\epsilon$ with $\epsilon \in (0,2)$. The overall predictions
are insensitive to the choice of a specific sector (for details,
see \cite{aps2,aps3}). States $\Psi (v,\phi)$ in $\H_\epsilon$ are
symmetric under the orientation inversion $\h{\Pi}$ and have
support on points $v= |\epsilon| + 4n$ where $n$ is an integer.
Wave functions $\Psi(v,\phi)$ in a generic sector solve
(\ref{hc3}) and are of positive frequency with respect to the
`internal time' $\phi$: they satisfy the `positive frequency'
square root
\be -i {\partial_\phi\,\Psi} = \sqrt{\Theta}\, \Psi\, .
\label{2}\ee
of Eq (\ref{hc3}). (The square-root is a well-defined (positive
self-adjoint) operator because $\Theta$ is positive and
self-adjoint.) The physical inner product is given by:
\be \ip{\Psi_1}{\Psi_2}\, = \, \sum_{v\in \{|\epsilon|+4n\}}
B(v)\, \bar\Psi_1(v, \phi_o) \Psi_2(v,\phi_o) \ee
and is `conserved', i.e., is independent of the `instant' $\phi_o$
chosen in its evaluation. On these states, the Dirac observables
act in the expected fashion:
\ba \h{p}_\phi \Psi &=& -i\hbar
{\partial_\phi\,\Psi}\nonumber\\
 \h{v}|_{\phi_o}\,\, \Psi (v,\phi) &=& e^{i
\sqrt{\Theta}(\phi-\phi_o)}\, v\, \Psi(v,\phi_o)\ea

What is the relation of this LQC description with the
Wheeler-DeWitt theory? It is straightforward to show \cite{aps3}
that, for $v \gg 1$, there is a precise sense in which the
difference operator $\Theta$ approaches the Wheeler DeWitt
differential operator $\ul{\Theta}$, given by
\be \label{wdw2} \ul\Theta \Psi(v,\phi) = {12\pi G}\,\, v\p_v
\big(v\p_v\Psi(v,\phi)\big) \ee
Thus, if one ignores the quantum geometry effects, Eq (\ref{hc3})
reduces to the Wheeler-DeWitt equation
\be \label{wdw1} \partial^2_\phi\Psi = -\ul{\Theta}\,\Psi. \ee
Note that the operator $\ul\Theta$ is positive definite and
self-adjoint on the Hilbert space $L^2_s(\R, \ub{B}(v)\dd v)$
where the subscript $s$ denotes the restriction to the symmetric
eigenspace of $\Pi$ and $\ub{B}(v) := Kv^{-1}$ is the limiting
form of $B(v)$ for large $v$. Its eigenfunctions $\ub{e}_k$ with
eigenvalue $\omega^2 (\ge 0)$ are 2-fold degenerate on this
Hilbert space. Therefore, they can be labelled by a real number
$k$:
\be \ub{e}_k(v) := \f{1}{\sqrt{2\pi}} \, e^{ik\ln |v|}\ee
where $k$ is related to $\omega$ via $\omega= \sqrt{12\pi G}|k|$.
They form an orthonormal basis on  $L^2_s(\R, \ub{B}(v)\dd v)$.  A
`general' positive frequency solution to (\ref{wdw1}) can be
written as
\be \label{wdwsol}\Psi(v, \phi) = \int_{-\infty}^{\infty}\, \dd k
\, \t\Psi(k)\, \ub{e}_k(v) e^{i\omega \phi} \ee
for suitably regular $\t\Psi(k)$. This expression will enable us
to show explicitly that the singularity is not resolved in the
Wheeler-DeWitt theory.

With the physical Hilbert space and a complete set of Dirac
observables at hand, we can now construct states which are
semi-classical at late times ---e.g., now--- and evolve them
`backward in time' numerically. There are three natural
constructions to implement this idea in detail, reflecting the
freedom in the notion of semi-classical states. In all cases, the
main results are the same \cite{aps2,aps3}. Here I will report on
the results obtained using the strategy that brings out the
contrast with the Wheeler DeWitt theory most sharply.
\begin{figure}
  \begin{center}
    \begin{minipage}{2.0in}
      \begin{center}
        \includegraphics[width=7cm,angle=20]{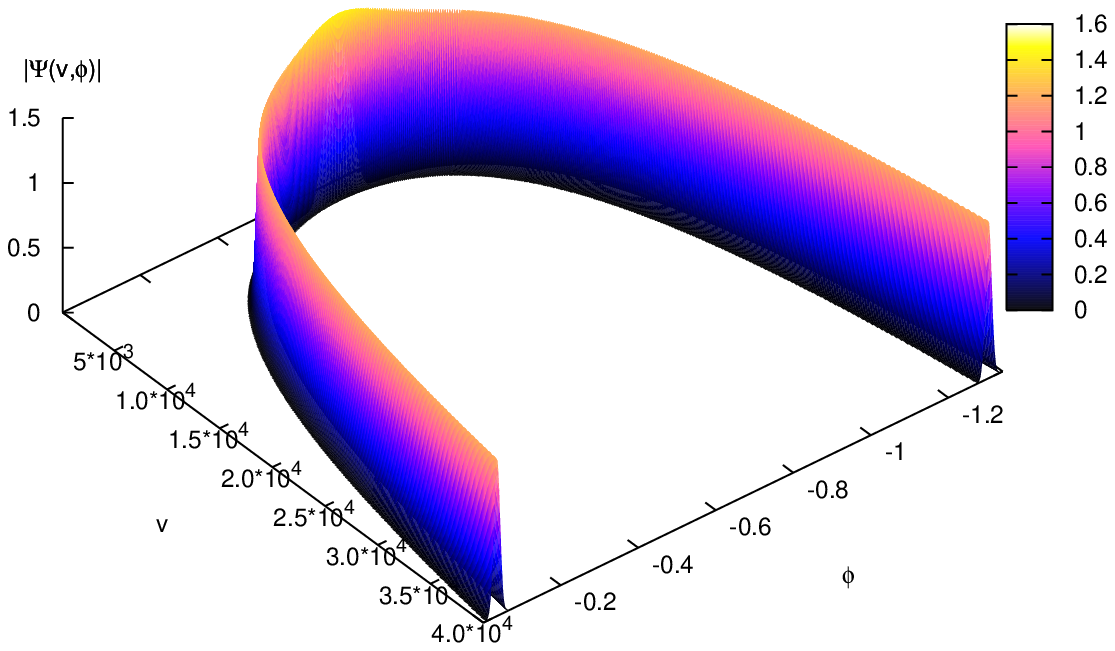}
      \end{center}
    \end{minipage}
    \hspace{2in}
    \begin{minipage}{2.0in}
      \begin{center}\small
        \includegraphics[width=5.5cm,angle=1]{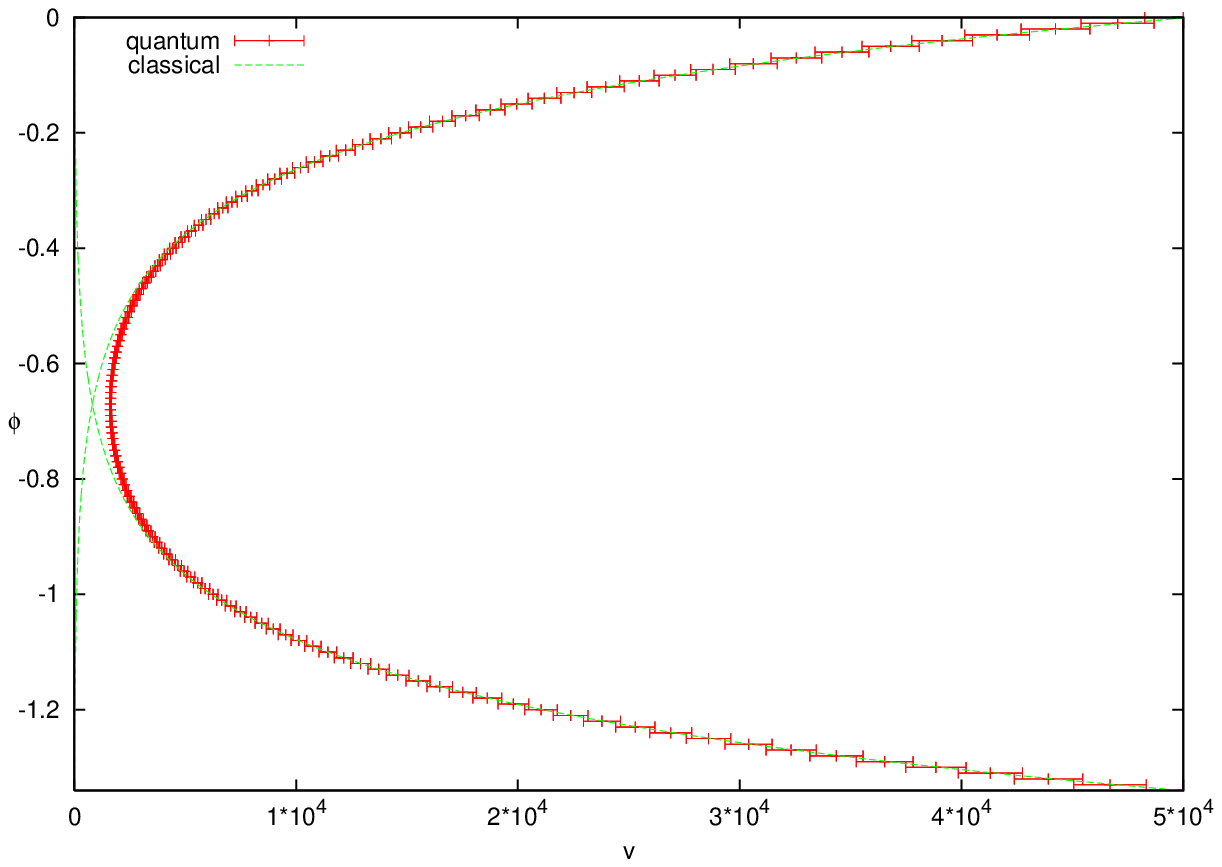}
      \end{center}
    \end{minipage}
    \caption{The figure on left shows the absolute value of the
wave function $\Psi$ as a function of $\phi$ and $v$. Being a
physical state, $\Psi$ is symmetric under $v \rightarrow -v$. The
figure on the right shows the expectation values of Dirac
observables $\h{v}|_{\phi}$ and their dispersions. They exhibit a
quantum bounce which joins the contracting and expanding classical
trajectories marked by fainter lines. In this simulation, the
parameters in the initial data are: $ v^\star = 5\times 10^4,\,\,
p_\phi^\star = 5\times 10^3 \sqrt{G}\hbar$ and  $ \Delta
p_\phi/p_\phi = 0.0025$.}
\end{center}
\end{figure}

As noted before, $p_\phi$ is a constant of motion. For the
semi-classical analysis, we are led to choose a large value
$p_\phi^\star$ ($\gg \sqrt{G}\hbar$). In the closed model, for
example, this condition is necessary to ensure that the universe
can expand out to a macroscopic size. Fix a point $(v^\star,
\phi_o)$ on the corresponding classical trajectory which starts
out at the big bang and then expands, choosing $v^\star \gg 1$. We
want to construct a state which is peaked at $(v^\star,
p_\phi^\star)$ at a `late initial time' $\phi\!=\!\phi_o$ and
follow its `evolution' backward. At `time' $\phi=\phi_o$, consider
then the function
\be \label{sc} \Psi(v, \phi_o) = \int_{-\infty}^\infty \dd k\,
\t\Psi(k)\, \ub{e}_k(v)\, e^{i\omega(\phi_o-\phi^\star)}, \quad
{\rm where}\,\,\, \t\Psi(k) = e^{-\f{(k-k^\star)^2}{2\sigma^2}}
\ee
where $k^\star = - p_\phi^\star/\sqrt{12\pi G\hbar^2}$ and
$\phi^\star = -\sqrt{1/12\pi G}\, \ln (v^\star) +\phi_o$. One can
easily evaluate the integral in the approximation  $|k^*| \gg 1$
and calculate mean values of the Dirac observables and their
fluctuations. One finds that, as required, the state is sharply
peaked at values $v^\star, p_\phi^\star$. The above construction
is closely related to that of coherent states in non-relativistic
quantum mechanics. The main difference is that the observables of
interest are not $v$ and its conjugate momentum but rather $v$ and
$p_\phi$ ---the momentum conjugate to `time', i.e., the analog of
the Hamiltonian in non-relativistic quantum mechanics. Now, one
can evolve this state backwards using the Wheeler-DeWitt equation
(\ref{wdw1}). It follows immediately from the form (\ref{wdwsol})
of the general solution to (\ref{wdw1}) and the fact that $p_\phi$
is large that this state would remain sharply peaked at the chosen
classical trajectory and simply follow it into the big-bang
singularity.

In LQC, we can use the restriction of (\ref{sc}) to points $v =
|\epsilon|+ 4n$ as the initial data and evolve it backwards
numerically. Now the evolution is qualitatively different (see
Fig.1). The state does remains sharply peaked at the classical
trajectory till the matter density reaches a critical value:
\be \rho_{\rm crit} =  \f{\sqrt{3}}{16\pi^2 \gamma^3 G^2 \hbar}\,
, \ee
which is about 0.82 times the Planck density. However, \emph{then
it bounces}. Rather than following the classical trajectory into
the singularity as in the Wheeler-DeWitt theory, the state `turns
around'. What is perhaps most surprising is that it again becomes
semi-classical and follows the `past' portion of a classical
trajectory, again with $p_\phi\! =\! p_\phi^\star$, which was
headed towards the big crunch. Let us we summarize the forward
evolution of the full quantum state. In the distant past, the
state is peaked on a classical, contracting pre-big-bang branch
which closely follows the evolution dictated by Friedmann
equations. But when the matter density reaches the Planck regime,
quantum geometry effects become significant. Interestingly, they
make gravity \emph{repulsive}, not only halting the collapse but
turning it around; the quantum state is again peaked on the
classical solution now representing the post-big-bang, expanding
universe. Since this behavior is so surprising, a very large
number of numerical simulations were performed to ensure that the
results are robust and not an artifact of the special choices of
initial data or of the numerical methods used to obtain the
solution \cite{aps2,aps3}.

For states which are semi-classical at late times, the numerical
evolution in exact LQC can be well-modelled by an effective,
modified Friedman equation :
\be \label{eff} \frac{\dot{a}^2}{a^2}\, =\, \frac{8\pi
G}{3}\,\,\rho\,\, \Big[1 - \frac{\rho}{\rho_{\rm crit}}\Big]
\ee
where, as usual, $a$ is the scale factor. In the limit $\hbar
\rightarrow 0$, $\rho_{\rm crit}$ diverges and we recover the
standard Friedmann equation. Thus the second term is a genuine
quantum correction. Eq. (\ref{eff}) can also be obtained
analytically from (\ref{hc3}) by a systematic procedure \cite{jw}.
But the approximations involved are valid only well outside the
Planck domain. It is therefore surprising that the bounce
predicted by the exact quantum equation (\ref{hc3}) is well
approximated by a naive extrapolation of (\ref{eff}) across the
Planck domain. While there is some understanding of this seemingly
`unreasonable success' of the effective equation (\ref{eff}),
further work is needed to fully understand this issue.

Finally let us return to the questions posed in the beginning of
this section. In the model, LQC has been able to answer all of
them. One can deduce from first principles that classical general
relativity is an excellent approximation till very early times,
including the onset of inflation in standard scenarios. Yet
quantum geometry effects have a profound, global effect on
evolution. In particular, the singularity is naturally resolved
without any external input and there is a classical space-time
also in the pre-big-bang branch. LQC provides a deterministic
evolution which joins the two branches.

\section{DISCUSSION}
\label{s4}

Even though there are several open issues in the formulation of
full quantum dynamics in LQG, detailed calculations in simple
models have provided hints about the general structure. It appears
that the most important non-perturbative effects arise from the
replacement of the local curvature term $F_{ab}^i$ by non-local
holonomies. This non-locality is likely to be a central feature of
the full LQG dynamics. In the cosmological model considered in
section \ref{s3}, it is this replacement of curvature by
holonomies that is responsible for the subtle but crucial
differences between LQC and the Wheeler-DeWitt theory.%
\footnote{Because early presentations emphasized the difference
between $B(v)$ of LQC and $\ub{B}(v) = Kv^{-1}$ of the
Wheeler-DeWitt theory, there is a misconception in some circles
that the difference in quantum dynamics is primarily due to the
non-trivial `inverse volume' operator of LQC. This is not correct.
In deed, in the model considered here, qualitative features of
quantum dynamics, including the bounce, remain unaffected if one
replaces by hand $B(v)$ with $\ub{B}(v)$ in the LQC evolution
equation (\ref{hc3}).}

By now a number of mini-superspaces and a few midi-superspaces
have been studied in varying degrees of detail. In all cases, the
classical, space-like singularities are resolved by quantum
geometry \emph{provided one treats the problem
non-perturbatively.} For example, in anisotropic mini-superspaces,
there is a qualitative difference between perturbative and
non-perturbative treatments: if anisotropies are treated as
perturbations of a background isotropic model, the big-bang
singularity is not resolved while if one treats the whole problem
non-perturbatively, it is \cite{mb-aniso}.

A qualitative picture that emerges is that the non-perturbative
quantum geometry corrections are \emph{`repulsive'}. While they
are negligible under normal conditions, they dominate when
curvature approaches the Planck scale and halt the collapse that
would classically have lead to a singularity. In this respect,
there is a curious similarity with the situation in the stellar
collapse where a new repulsive force comes into play when the core
approaches a critical density, halting further collapse and
leading to stable white dwarfs and neutron stars. This force, with
its origin in the Fermi-Dirac statistics, is \emph{associated with
the quantum nature of matter}. However, if the total mass of the
star is larger than, say, $5$ solar masses, classical gravity
overwhelms this force. The suggestion from LQC is that, a new
repulsive force \emph{associated with the quantum nature of
geometry} may come into play near Planck density, strong enough to
prevent the formation of singularities irrespective of how large
the mass is. Since this force is negligible until one enters the
Planck regime, predictions of classical relativity on the
formation of trapped surfaces, dynamical and isolated horizons
would still hold. But assumptions of the standard singularity
theorems would be violated. There would be no singularities, no
abrupt end to space-time where physics stops. Non-perturbative,
background independent quantum physics would continue.

Returning to the issue of the Beginning, the big-bang in
particular appears to be an artifact of the assumption that the
continuum, classical space-time of general relativity should hold
at all scales. LQC strongly suggests that this approximation
breaks down when the matter reaches Planck density. One might have
at first thought that, since this is a tiny portion of space-time,
whatever quantum effects there may be, they would have negligible
effect on global properties of space-time and hence almost no
bearing on the issue of The Beginning. However, detailed LQC
calculations have shown that this intuition may be too naive. The
`tiny portion' may actually be a bridge to another large universe.
The physical, quantum space-time of could be significantly larger
than what general relativity had us believe. The outstanding open
issue is whether this scenario persists when inhomogeneities are
adequately incorporated in the analysis.

\bigskip
\textbf{Acknowledgments:} I would like to thank Martin Bojowald,
Jerzy Lewandowski, and especially Tomasz Pawlowski and Parampreet
Singh for collaboration and numerous discussions. This work was
supported in part by the NSF grants PHY-0354932 and PHY-0456913,
the Alexander von Humboldt Foundation, the Krammers Chair program
of the University of Utrecht and the Eberly research funds of Penn
State.

\end{document}